\documentclass{elsart}
\usepackage{graphicx}
\begin{document}

\newcommand{\var}{\mbox{Var}}
\newcommand{\average}[1]{E\left[#1\right]}

\begin{frontmatter}

\title{Downside Risk analysis applied to the Hedge Funds universe}
\author{Josep Perell\'o}
\ead[url]{http://www.ffn.ub.es/perello}
\ead{josep.perello@ub.edu}
\address{Departament de F\'{\i}sica Fonamental, 
Universitat de Barcelona, Diagonal, 647, 08028-Barcelona, Spain}
%\date{\today}
\maketitle

\begin{abstract}
Hedge Funds are considered as one of the portfolio management sectors which shows a fastest growing for the past decade. An optimal Hedge Fund management requires an appropriate risk metrics. The classic CAPM theory and its Ratio Sharpe fail to capture some crucial aspects due to the strong non-Gaussian character of Hedge Funds statistics. A possible way out to this problem while keeping the CAPM simplicity is the so-called Downside Risk analysis. One important benefit lies in distinguishing between good and bad returns, that is: returns greater or lower than investor's goal. We revisit most popular Downside Risk indicators and provide new analytical results on them. We compute these measures by taking the Credit Suisse/Tremont Investable Hedge Fund Index Data and with the Gaussian case as a benchmark. In this way an unusual transversal lecture of the existing Downside Risk measures is provided.
\end{abstract}
\end{frontmatter}

\section{Introduction}

Hedge Funds are considered as one of the portfolio management sectors which shows a fastest growing for the past few years~\cite{lhabitant,seco}. These funds have been in existence for several decades but they do not have become popular until the 1990's. It is said that the Hedge Funds are capable of making huge profits but sometimes also suffer spectacular losses. Due to their at least apparent high and unpredictable fluctuations, it is necessary to keep the risks we take when we trade with Hedge Funds under rigorous control~\cite{till2,till}.

The Capital Asset Pricing Model (CAPM) is the classic method for quantifying the risk of a certain portfolio~\cite{markowitz,sharpe}. Basically, the so-called Ratio Sharpe~\cite{sharpe} evaluates the quality of a certain asset by normalizing the asset growth expectation with its volatility, that is its standard deviation. Thus, based on the fact that the asset growth expectation must be large and volatility low, a {\it good} Hedge Fund holds a high Ratio Sharpe. And the better the Hedge Fund the more attractive and advisable is to invest in this fund. Typically, Hedge Fund managers begin to trade with an specific Hedge Fund only when this fund gets an annual Ratio Sharpe greater than 1~\cite{lhabitant}. It is considered that only if Ratio Sharpe crosses this threshold the fund can provide benefits after the trading costs removal.

However, the CAPM theory is sustained under the hypothesis that underlying asset is Gaussian distributed. In this case, investor only needs to know the mean and the variance of the return. As it has been observed~\cite{mantegna,mmp,perello,perello2}, this appears to be an unrealistic scenario in financial markets with important implications in the risk analysis within the mean-variance framework (see for instance~\cite{estrada}). The situation is much more dramatic in the Hedge Fund universe since these funds are clearly non-Gaussian having wild fluctuations and strong asymmetries in price changes. These funds are characterized by their big sensitivity to the market crashes and by trading with products such as derivatives that show a pronounced skewness in their price distribution. For instance, a very well-known commodity trader adviser (CTA) Hedge Fund had a poor Ratio Sharpe (0.19) but, despite this mediocre mark, their earnings during the 2000 raised beyond the 40\%~\cite{lhabitant}. Conversely, after 31 months of trading, the famous fund Long-Term Capital Management (LTCM) had an appealing ratio (4.35) and nothing seemed to forecast its posterior debacle~\cite{lhabitant}. These two examples are not exceptional cases, and they make us reexamine the validity of the CAPM theory. There is evidence that the CAPM method is not complete enough for evaluating the risks involved in the Hedge Fund management. 

Our aim here is to explore some alternatives in the context of the so-called Downside Risk analysis~\cite{lhabitant,estrada,sortino3}. The main contributions of the current paper are the following. We introduce some of the most popular Downside Risk indicators sparse in the literature. We also provide new analytical results related to these risk measures and finally make a large set of empirical measurements to the Credit Suisse/Tremont Investable Hedge Fund Index Data. The application of all these Downside Risk indicators to the same data set has thus allowed to revisit them in a transversal way which is quite unusual in the literature. We have finally found that representing the indicators in terms of a modified Ratio Sharpe (see below) is more appropriate than doing it as a function of the investor's goal. This replacement with respect what is typically done in the literature makes possible a better data collapse and it in turn enables to compare easily the performance between different Hedge Fund trading styles.

The paper is structured as follows. Section~\ref{data} briefly describes the data set used for the Downside Risk indicators. The following section is devoted to present the backgrounds of the Downside Risk approach. Afterwards, we present the Adjusted Ratio Sharpe in Section~\ref{adjusted}, the Sortino ratios in Section~\ref{sortino}, and the Gain-Loss Ratio is left to Section~\ref{g-lsec}. The equivalence between the Omega function and the Gain-Loss Ratio under specific circumstances is given in Section~\ref{omega-sec} while a discussion about the error behind the risk measures we use is left to Section~\ref{errors}.
Finally, Section~\ref{conc} provides some conclusions.

\section{The Hedge Fund data set\label{data}}

There are several third-party agencies that collect and distribute data in Hedge Fund performance~\cite{lhabitant}. For this paper, we have used the data supplied by the Credit Suisse/Tremont (CST) Index~\cite{web}. This company is a joint venture between Credit Suisse and Tremont Advisers Inc which combines their resources to set several benchmarks for a large collection of Hedge Fund strategies. They provide a master index and series of sub-indices that represent the historical returns for different Hedge Fund trading styles~\cite{lhabitant,web}.

The weight of each fund in an index is given by the relative size of its assets under management. This makes the CST Index the first asset-weighted indices in the industry. Asset-weighting, as opposed to equal-weighting, provides a more accurate depiction of an investment in the asset class. In addition, CST has a web site~\cite{web} that provides an up-to-date and historical data and allows the user to select and download data. Information available is public. The selection of funds for the CST indices is done every quarter. The process starts by considering all 2,600 US and offshore Hedge Funds contained in the TASS database, with the exception of funds of funds and managed accounts.

In the present case, we have analyzed the monthly data for these indices during the period between 31st December of 1993 until the 31st January of 2006. This period corresponds to 145 data points for each Hedge Fund style. This is not a very large data set but it is enough to perform a reasonably fair and reliable statistical estimation of the quantities we here deal with. 

\begin{figure}
\begin{center}
\includegraphics{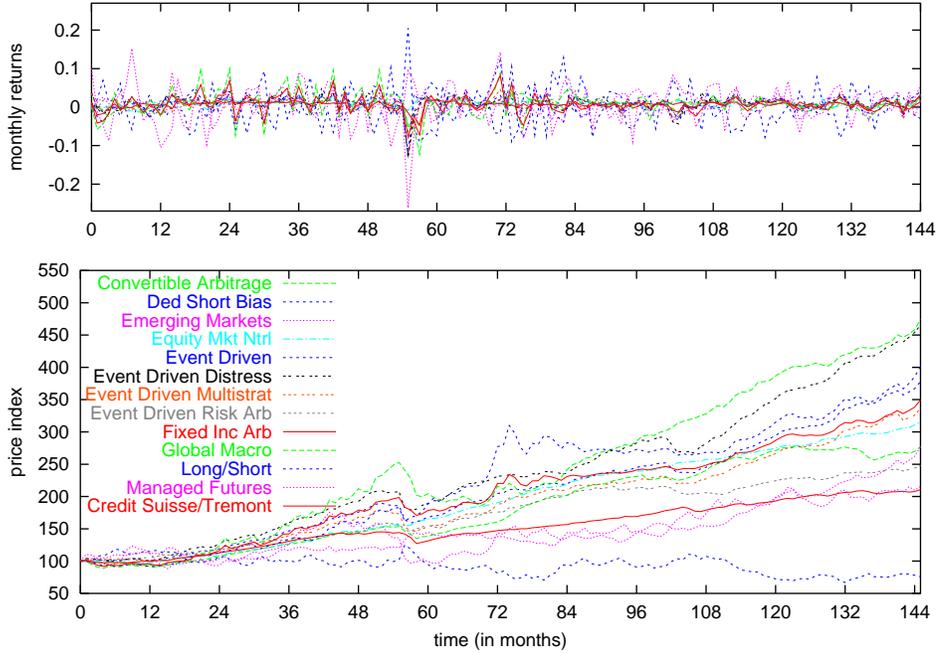}
\end{center}
\caption{The price and the monthly return change time series for the Credit Suisse/Tremont (CST) Index and subindices during the period between 31st December of 1993 until the 31st January of 2006.}
\label{prices}
\end{figure}

In Fig.~\ref{prices} we show the indices dynamics that were all normalized to 100 at the beginning of 1994. We also show the monthly logarithmic return change $R(t)=\ln[S(t+\Delta)/S(t)]$ where $S(t)$ is current price index and $\Delta$ is fixed and equals to one month. In what follows the return change is always over one month and this is the reason why we avoid to specify the value $\Delta$. Table~\ref{tab} shows how the mean-variance framework fails to explain the statistics of the majority of Hedge Fund styles monthly returns. The kurtosis can raise to values larger than 20 while the skewness is usually negative and may take values larger than 3.

\begin{table}
\caption{Main statistical values for the whole set of Hedge Fund style indices during the period between 31st December of 1993 until the 31st January of 2006. We show the first moment, the standard deviation, the kurtosis and the skewness for the monthly returns. Most of the indices have a kurtosis larger than one and some of them also have a non negligible negative skewness. The mean-variance framework might fit well only for very few of them (the ``Managed Futures" and the ``Equity Market Neutral" styles).}
\bigskip
\begin{center}
\begin{tabular}{lrrrr}
\hline
\hline
Hedge Funds indices & mean & std dev & kurt & skew \\\hline
Credit Suisse/Tremont Index & $0.009\pm 0.002$ & $0.0225\pm 0.0010$ & 2.3 & -0.040\\
Convertible Arbitrage & $0.0070\pm 0.0011$ & $0.0138\pm 0.0004$ &3.2 & -1.4\\
Dedicated Short Bias & $-0.002\pm 0.004$ & $0.049\pm 0.004$ & 1.2& 0.63\\
Emerging Markets & $0.007\pm 0.004$ & $0.048\pm 0.007$ & 6.5& -1.1\\
Equity Market Neutral & $0.0079\pm 0.0007$  & $0.00838\pm 0.00011 $ & 0.34& 0.30\\
Event Driven & $0.0092\pm 0.0014$ & $0.017\pm 0.002$ & 28 & -3.8\\
Fixed Income Arbitrage & $0.0051\pm 0.0009$ & $0.0109\pm 0.0005$ & 18& -3.2\\
Global Macro & $0.011\pm 0.003$ & $0.032\pm 0.002$ & 3.0& -0.21\\
Long/Short & $0.010\pm0.002$ & $0.029\pm 0.002$ & 4.0& -0.042\\
Managed Futures & $0.005\pm 0.003$ & $0.034\pm 0.002 $ & 0.47& -0.099\\
Event Driven Distressed & $0.011\pm 0.002$ & $0.019\pm 0.002$ & 22& -3.3\\
Event Driven Multistrategy & $0.008\pm 0.002$ & $0.018\pm 0.002$ & 19& -2.9\\
Event Driven Risk Arbitrage & $0.0063\pm 0.0010$ & $0.0121\pm 0.0004$ & 7.2& -1.4\\
\hline
\hline
\bigskip
\end{tabular}
\end{center}
\label{tab}
\end{table}

\section{The Downside Risk Metrics: Main definitions\label{dr}}

For the reasons mentioned above, the so-called Downside Risk analysis has been gaining wide acceptance in recent years~\cite{till2,till,estrada}. One important benefit of the Downside Risk lies in distinguishing between {\it good} and {\it bad} returns: Good returns are greater than the goal, while bad returns are the ones below the goal. The Downside Risk measures incorporate an investor's goal explicitly and define risk as not achieving the goal. In this way, the further below the goal, the greater the risk. And, in the opposite side, returns over the goal does not imply any risk. Within this approach, a portfolio's riskiness may be perceived differently by investors with different goals. This is more realistic than the CAPM theory approach where all investors have the same risk perception. 

Typically, the target return is set to be to the minimum acceptable return for considering profitable the trading operation. And the statistical risk would be then associated with the unsuccessful tentatives of obtaining a higher return than the target return. However, the target return could also be related to the maximum loss that a Hedge Fund can afford measuring risk in a somewhat similar way as the Value at Risk measures do~\cite{wilmott}. We will here cover a broad window ranging the annual target return values from -30\% to +30\%.

We consider the price of the asset $S$ at time $t$ and its initial price $S_0$ at time $t=0$. Let us thus define the Excess Downside as
\begin{equation}
D(R,T)\equiv\max[0,T-R]=
\left\{
\begin{array}{rr}
T-R & \mbox{if } T>R, \\
0 & \mbox{if } T\leq R;
\end{array}
\right.
\label{D}
\end{equation}
where $R\equiv\ln(S/S_0)$ is the subsequent monthly return change and $T$ is the target return. Observe that the mathematical expression for the Excess Downside is identical to the one for the payoff of the European put option~\cite{wilmott}. 

We can ineed study the first and second moments of the Excess Downside $D(T)$. Recall that the first moment is defined as
\begin{equation}
\mu_-(T)\equiv\average{D(R,T)}=\int_{-\infty}^{T} (T-R) p(R) dR,
\label{D1}
\end{equation}
while the second moment reads
\begin{equation}
d(T)^2\equiv\average{D(R,T)^2}=\int_{-\infty}^{T} (T-R)^2 p(R) dR,
\label{D2}
\end{equation}
where $p(R)$ is the probability density function (pdf) of the return $R$. The square root of the second moment~(\ref{D2}) is also called Excess Downside Deviation (EDD). 

Figure~\ref{adj-fig} represents the empirical ratio $d(T)/\sigma$ between the EDD and the volatility (standard deviation of the returns) in terms of the target return $T$ with $\lambda=(\mu-T)/\sigma$. The other two parameters involved are
\begin{equation}
\mu=\average{R},
\label{mu}
\end{equation}
which is the first moment of the return and 
\begin{equation}
\sigma^2=\var[R]
\label{s}
\end{equation} 
is the return variance. Both are directly computed from historical data. Observe that the ratio $d/\sigma$ for the Hedge Fund data indices may differ significantly from the Gaussian case, specially for $T$'s lower than $\mu$ ($\lambda>0$). We discuss deeper the results of this plot in the next section. 

However, before proceeding we shall notice that the CST Index results have been plotted with error bars for the y-axis. The bars represent the standard error of the measurements. Figure~\ref{adj-fig} plots an algebraic combination of averages and we apply standard rules of error propagation of each standard error measurement. We have done it done for the rest of trading style indices with very similar conclusions but in order to lighten the information in the plot we do not show these errors bars. We neither show the standard error for the x-axis. Their error bars do not change for different values of $\lambda$ and they typically affect the second significative digit of lambda. All these comments are also valid for the rest of figures and empirical results in the paper. We will go a bit further on this topic in Section~\ref{errors}.

\section{The Adjusted Ratio Sharpe\label{adjusted}}

A first possible extension of the method aims to keep the CAPM approach but with a rough correction based on the empirical EDD computed as defined in Eq.~(\ref{D2}). This is probably the simplest sophistication to the mean-variance framework. Its interest is based on the fact that it is aimed to replace the volatility $\sigma$ by a more appropriate risk measure such as $d(T)$. We recall that the CAPM measures risk of a certain asset with the well-known
\begin{equation}
\mbox{Ratio Sharpe}=\frac{\mu-r}{\sigma},
\label{sharpe}
\end{equation}
where $r$ is the risk-free interest rate ratio, $\mu=\average{R}$ and $\sigma^2=\var[R]$. Johnson etal.~\cite{johnson} propose an Adjusted Ratio Sharpe as ``the Ratio Sharpe that would be implied by the fund's observed Downside Deviation if returns were distributed normally". 

Let us study further this risk measure both analytically and empirically. We first assume that the returns are Gaussian:
\begin{equation}
p(R)=\frac{1}{\sqrt{2\pi}\sigma}\exp\left[-\frac{(R-\mu)^2}{2\sigma^2}\right].
\label{pr}
\end{equation}
We also define a modified Ratio Sharpe with the quotient
\begin{equation}
\lambda\equiv\frac{\mu-T}{\sigma}.
\label{lambda}
\end{equation}
This variable $\lambda$ is very important not only in this rough correction of the Ratio Sharpe but also in the analysis we will perform for other alternative risk indicators we will show herein. The effort to represent the risk measures in terms of $\lambda$ is of the new contributions of this paper. The forthcoming measures in the next sections can be all represented exclusively in terms of lambda if the underlying is Gaussian distributed. Nonetheless, it will be also helpful to keep on working with the $\lambda$ even for the empirical data set where Gaussian hypothesis is weakly sustained.

\begin{figure}
\begin{center}
\includegraphics{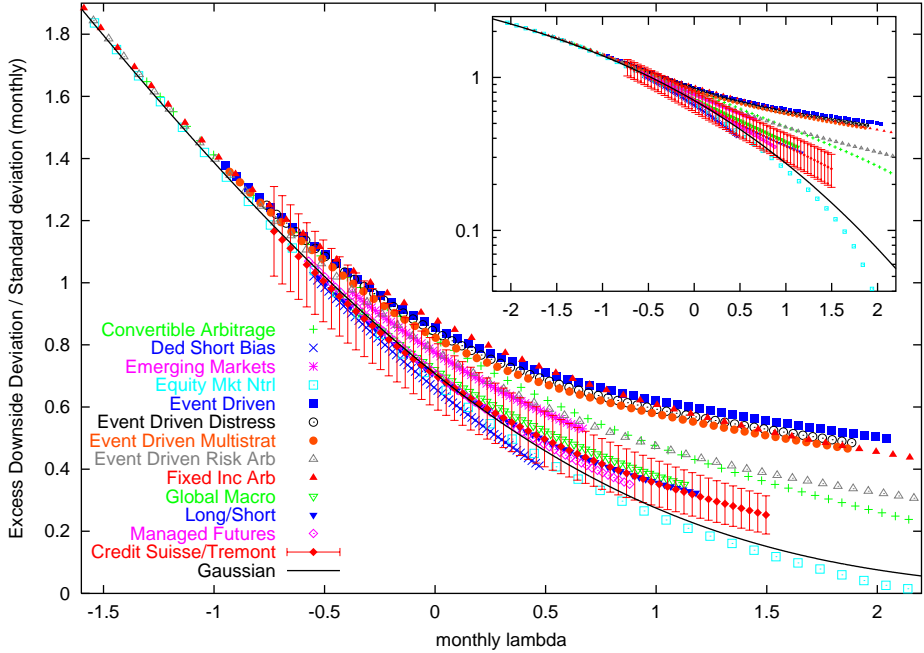}
\end{center}
\caption{The quotient between the Excess Downside Deviation and the volatility for the CST Index and subindices for the period between the 31st Dec of 1993 until the 31st Jan of 2006. We show the quotient $d(T)/\sigma$ in terms of $\lambda=(\mu-T)/\sigma$ for several Hedge Fund styles when target return $T$ is between $-30\%$ and $30\%$ annual rates. The inset provides the same results but in a logarithmic scale. The empirical results are compared with the Gaussian case given by Eq.~(\ref{lambda-eq}). The error bars in the CST Index correspond to the standard errors of each $\lambda$ measurement.}
\label{adj-fig}
\end{figure}

Therefore, from the Excess Downside Deviation~(\ref{D2}) and assuming Gaussian returns we can write
\begin{equation}
\frac{d(T)^2}{\sigma^2}=\left(\lambda^2+1\right)N(-\lambda)-\lambda N'(\lambda).
\label{lambda-eq}
\end{equation}
where $N(\cdot)$ is the probability function
\begin{equation}
N(\lambda)=\frac{1}{\sqrt{2\pi}}\int_{-\infty}^\lambda\exp\left(-\frac{y^2}{2}\right)dy,
\label{No}
\end{equation}
while its derivative is denoted by a prime and reads $\exp\left(-\lambda^2/2\right)/\sqrt{2\pi}$. 

Figure~\ref{adj-fig} plots the $d(T)/\sigma$ ratio in terms of the modified Ratio Sharpe~(\ref{lambda}) for a broad range of target returns $T$, from -30\% until 30\% annualized rates. We represent data in terms of $\lambda$ by taking the empirical averages $\mu$ and $\sigma$ (cf. Eqs.~(\ref{mu})-(\ref{s})). When $\lambda<0$ ($T>\mu$), we can observe that the empirical $d(T)/\sigma$ follows reasonably well the Gaussian curve given by Eq.~(\ref{lambda-eq}). However, when $\lambda>0$ ($T<\mu$), the results may differ significantly from the Gaussian curve. These more pronounced differences for $T<\mu$ can be adduced to the effect of the negative skewness. In this case, we are mainly taking the left wing of the distribution where returns are smaller than the average. The more negative the skewness is, the larger the $d(T)/\sigma$ ratio. For instance, the ``Event Driven" index has a skewness more negative than -3 and it describes one of the curves with higher values of the $d(T)/\sigma$ ratio. In contrast, the ``Equity Market Neutral" index remains even below the Gaussian curve mainly because of its slighlty positive skewness.

We can finally numerically invert the ratio $d(T)/\sigma$ obtained from Eq.~(\ref{lambda-eq}). We will thus obtain the so-called Adjusted Ratio Sharpe. That is: the lambda that corresponds to the Excess Downside Deviation ratio in the event of returns are Gaussian and then accomplishing the equality~(\ref{lambda-eq}). We show the resulting empirical results in Table~\ref{emp-ad} for the special case when $T=5\%$. The Adjusted Ratio Sharpe may also differ significantly from the Ratio Sharpe placed far outside its standard error region.

\begin{table}[tbp]
\caption{The monthly Adjusted Ratio Sharpe for several Hedge Fund indices. We first derive the monthly Excess Downside Deviation ratio $d(T)/\sigma$ for several Hedge Fund indices when annual return growth is $T=5\%$. Once we get these quantities we numerically invert Eq.~(\ref{lambda-eq}). We thus compare these results with the Ratio Sharpe which is takes as a risk-free interest ratio the data from Bloomberg with ticker US0001M Index which corresponds to the one-month LIBOR index (London Interbank Offered British Bankers Association Rate).}
\begin{center}
\begin{tabular}{lrrr}
\hline
\hline
Hedge Funds & $d(T)/\sigma$ & Adj. Ratio Sharpe & Ratio Sharpe \\\hline
Credit Suisse/Tremont Index & 0.609 & 0.63 & $-0.18\pm 0.06$\\
Convertible Arbitrage & 0.729 & -0.13 & $0.29\pm 0.07$\\
Dedicated Short Bias & 0.735 & -0.16 & $-0.07\pm 0.16$\\
Emerging Markets & 0.746 & -0.23 & $-0.002\pm 0.002$\\
Equity Market Neutral & 0.447 & 1.82 & $0.55\pm 0.08$\\
Event Driven & 0.756 & -0.29 & $0.25\pm 0.09$\\
Fixed Income Arbitrage & 0.845 & -0.80 & $0.11\pm 0.04$\\
Global Macro & 0.626 & 0.51 & $0.22\pm 0.07$ \\
Long/Short & 0.612 & 0.61 & $0.17\pm 0.06$\\
Managed Futures & 0.688 & 0.12 & $0.03\pm 0.02$\\
Event Driven Distressed & 0.725 & -0.11 & $0.29\pm 0.10$\\
Event Driven Multistrategy & 0.740 & -0.20 & $0.19\pm 0.07$\\
Event Driven Risk Arbitrage & 0.705 & 0.01 & $0.21\pm 0.06$\\
\hline
\hline
\end{tabular}
\bigskip
\end{center}
\label{emp-ad}
\end{table}

\section{The Sortino and Upside Potential ratios\label{sortino}}

\begin{figure}
\begin{center}
\includegraphics{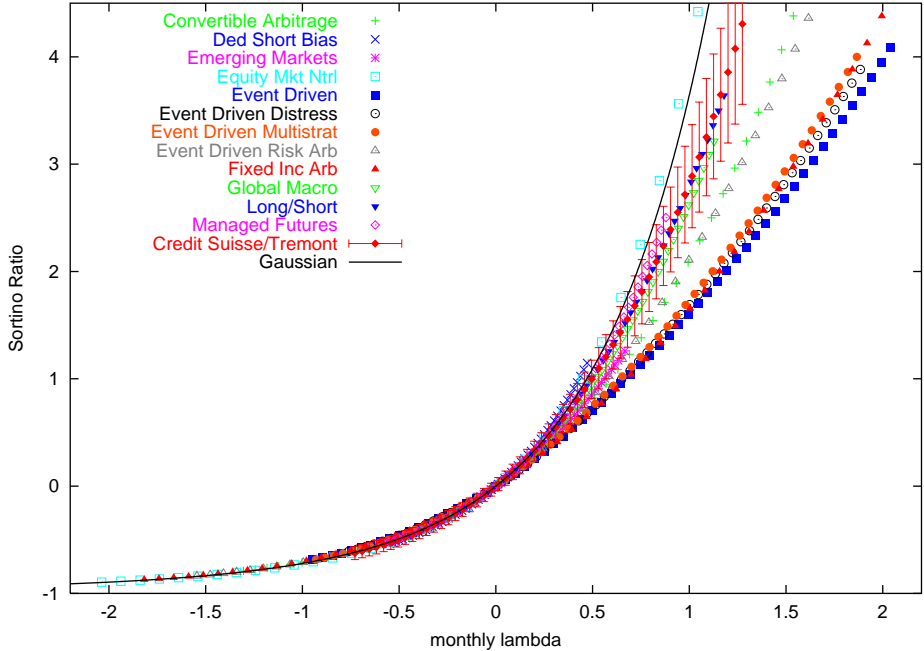}
\end{center}
\caption{The monthly Sortino Ratio for the Credit Suisse/Tremont (CST) Index and subindices during the period between 31st December of 1993 until the 31st January of 2006. We show the monthly SoR$(T)$ given by Eq.~(\ref{SoR}) in terms of lambda given by Eq.~(\ref{lambda}) for several Hedge Fund styles when target return is between $T=-30\%$ and $T=30\%$ annual rates. We compare them with the Gaussian Sortino Ratio~(\ref{SoRG}) and observe that the historical data results are not very far from the Gaussian hypothesis. The error bars in the CST Index correspond to the standard error for each $\lambda$ measurement.}
\label{sortino-fig}
\end{figure}

\begin{figure}
\begin{center}
\includegraphics{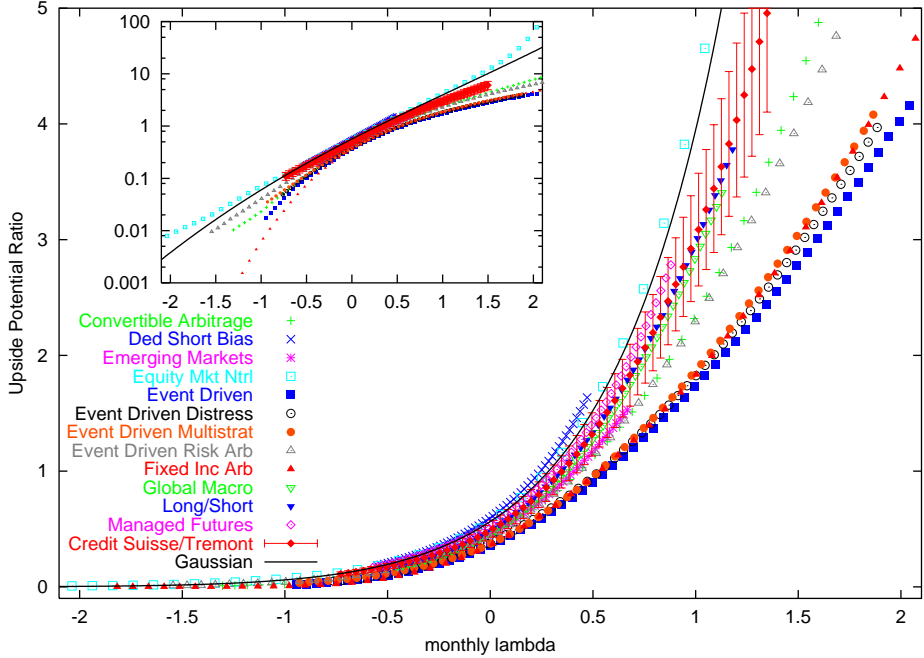}
\end{center}
\caption{The monthly Upside Potential Ratio for the Credit Suisse/Tremont (CST) Index and subindices during the period between 31st December of 1993 until the 31st January of 2006. We show the UPR$(T)$ given by Eq.~(\ref{SoR}) in terms of $\lambda$ given by Eq.~(\ref{lambda}) for several Hedge Fund styles and when target return is between $T=-30\%$ and $T=30\%$ annual rates. We compare them with the Gaussian UPR~(\ref{SoR2G}). The inset shows the same plot but with a logarithmic scale in the UPR axe. The error bars in the CST Index correspond to the standard errors of the quotient for each $\lambda$ measurement.}
\label{upr-fig}
\end{figure}

Sortino etal. propose in their works~\cite{sortino0,sortino1,sortino2,pedersen} more drastic modifications of the Ratio Sharpe. In contrast with the Adjusted Ratio Sharpe, the pair of indicators we here introduce are self-consistent measures that do not need to assume that the underlying asset is Gaussian. All averages involved in these indicators are directly computed from the historical data. 

However, these two new ratios have a similar apperance than the Ratio Sharpe. In both cases, the risk-free rate is replaced by the target return and the volatility by the Excess Downside Deviation. We here also claim that the modified Ratio Sharpe $\lambda$ will still be useful to represent all Hedge Fund styles into a single plot.

The first tentative is the so-called Sortino Ratio (SoR) defined as follows:
\begin{equation}
\mbox{SoR}(T)=\frac{\mu-T}{d(T)},
\label{SoR}
\end{equation}
where $d(T)$ is the EDD given by Eq.~(\ref{D2}). There is a sophistication made by the same Sortino whose computation also replaces return average $\mu$ by the excess return average. This new measure is worried about the return average up to the target return. Thus, the Upside Potential Ratio (UPR) is defined as
\begin{equation}
\mbox{UPR}(T)=\frac{\mu_+(T)}{d(T)},
\label{SoR2}
\end{equation}
where
\begin{equation}
\mu_+(T)=\int_T^\infty (R-T) p(R)dR,
\label{mu+}
\end{equation}
or equivalently $\mu_+(T)=\mu-T+\mu_-(T)$ (cf. Eq.(\ref{D1})). It is indeed possible to write one risk measure in terms of the other. That is: $\mbox{UPR}(T)=\mbox{SoR}(T)+\mu_-(T)/d(T)$.

Both measures behave as the Ratio Sharpe. The greater ratio corresponds to the better asset. Let us calculate the quotients given Eqs.~(\ref{SoR}) and~(\ref{SoR2}) in case that returns were Gaussian and here comes our new contribution on the study of the SoR and UPR measures. We first need the Gaussian $d(T)$ which is already given by Eq.~(\ref{lambda-eq}). In such a case the Sortino Ratio reads
\begin{equation}
\mbox{SoR}(T)=\frac{\lambda}{\left[\left(\lambda^2+1\right)N(-\lambda)-\lambda N'(\lambda)\right]^{1/2}},
\label{SoRG}
\end{equation}
while, again taking into account that under the Gaussian hypothesis the upside average~(\ref{mu+}) is
\begin{equation}
\mu_+(T)=\sigma\lambda N\left(\lambda\right)+\sigma N'(\lambda),
\label{mu+G}
\end{equation}
the Upside Potential Ratio reads
\begin{equation}
\mbox{UPR}(T)=\frac{\lambda N\left(\lambda\right)+N'(\lambda)}{\left[\left(\lambda^2+1\right)N(-\lambda)-\lambda N'(\lambda)\right]^{1/2}}.
\label{SoR2G}
\end{equation}
We note that our results on both risk measures have been expressed in terms of the modified Ratio Sharpe
$\lambda=(\mu-T)/\sigma$. And we here claim that this is still convenient even in the case we do not have a Gaussian distribution for the returns. 

Some special and limiting cases are
$$
-1<\mbox{SoR}(T)<\infty \qquad \mbox{and} \qquad 0<\mbox{UPR}(T)<\infty,
$$
where upper and lower bounds respectively correspond to the limiting cases $T\rightarrow \infty$ and $T\rightarrow -\infty$. 

These measures could be annualized as it was done with the Ratio Sharpe. Recall that the monthly Ratio Sharpe is annualized when we multiply the ratio by the factor $\sqrt{12}$. However, in principle, this is not as easy in the present case as it has been thoroughly investigated in Ref.~\cite{sortino1}.

We represent the Sortino and Upside Potential ratios in Figs.~\ref{sortino-fig} and~\ref{upr-fig}. We there plot the Gaussian case as a benchmark and in terms of lambda. The empirical data is also shown in terms of $\lambda$ to get the results comparable. For the empirical data we have computed the risk measures for a broad range of target returns $T$, from -30\% until 30\% annualized rates. At first sight, we do not perceive much differences between the two plots. We may however say that UPR is able to scatter in a slightly better way the different trading styles. In both risk measures and in the linear scale, the differences with respect the Gaussian curve become important only for positive $\lambda$ ($\mu>T$). However, the UPR risk measure is the only that can be also represented in a logarithmic scale. The relative distances to the Gaussian curve are quite symmetric for negative and positive $\lambda$'s. For this reason, we consider more powerful the UPR. In addition and from the error bars, we can also state that the replacement of the mean by the upside average does not bring much more noise to the UPR risk measure in comparison to the SoR risk measure noise. In general, we can also say that the large errors bars does not allow to get reliable conclusions for positive and moderate values of lambda.

In both risk measures and for a broad range of lambda, the index style closer to the Gaussian curve corresponds to the ``Equity Market Neutral". The set of trading styles could be sorted in several groups and without observing big discrepancies in that classification depending on which risk measure we take. The resulting groups are also consistent with the ones we can identify in Fig.~\ref{adj-fig}. Particularly, for large $\lambda$ (negative $T$) we can easily see that the risk in most of the ``Event Driven" indices and the ``Fixed Income Arbitrage" index is comparable. The reason why Gaussian curve is beyond most of the indices risk measures for positive lambda should be found in the negative and nonnegligible skewness in the data set of these Hedge Fund indices.

\section{The Gain-Loss Ratio\label{g-lsec}}

\begin{figure}
\begin{center}
\includegraphics{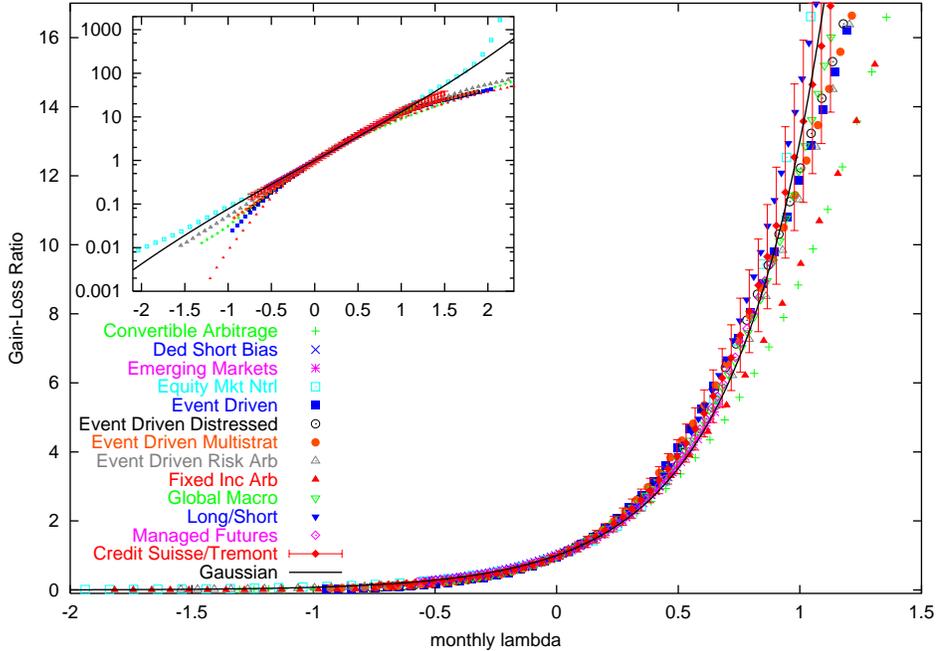}
\end{center}
\caption{The monthly Gain-Loss Ratio for the Credit Suisse/Tremont (CST) Index and subindices during the period between 31st December of 1993 until the 31st January of 2006. We show the ratio given by Eq.~(\ref{gl}) in terms of lambda given by Eq.~(\ref{lambda}) for several Hedge Fund styles and when target return is between $T=-30\%$ and $T=30\%$ annual rates. We compare them with the Gaussian case~(\ref{glG}). The inset shows the same plot but with a logarithmic scale in the GL axis. The error bars in the CST Index correspond to the standard errors of each $\lambda$ measurement.}
\label{gl-fig}
\end{figure}

Bernardo and Ledoit~\cite{bernardo} propose another risk measure called Gain-Loss Ratio (GL). This mesaure is probably the most well-grounded measure between the existing alternatives to the CAPM theory. In contrast with the Sortino ratios, GL have no comparable magnitudes with the Ratio Sharpe.

In the simplest scenario, the attractiveness of an investment opportunity is measured by the Gain-Loss Ratio
\begin{equation}
\mbox{GL}(T)=\frac{\mu_+(T)}{\mu_-(T)},
\label{gl}
\end{equation}
which is the fraction between the averages of positive and negative parts of the payoff after removing the trading costs included in the target return $T$. 

The framework provides an alternative approach to ``asset pricing in incomplete markets that bridges the gap between the two fundamental approaches in finance: model-based pricing and pricing by no arbitrage"~\cite{bernardo}. The GL$(T)$ ratio constitutes the basis to an alternative asset pricing framework. By limiting the maximum Gain-Loss ratio, the admissible set of pricing kernels can be restricted and it is also possible also to constrain the set of prices that can be assigned to assets~\cite{bernardo}. In other words, we admit that there are arbitrage opportunities but limited in a certain range of prices. In the same way, Bernardo and Ledoit~\cite{bernardo} state that the theoretical no arbitrage assumption is related to the mathematical demand that the GL is 1. 

We can now explore this risk measure in a similar way as done in the previous sections and this corresponds to our new contribution to this risk indicator. Following the notation used above we have that
\begin{equation}
\mu_-(T)\equiv\average{(T-R)^+}=\int_{-\infty}^T (T-R) p(R)dR,
\label{mu-def}
\end{equation}
and
\begin{equation}
\mu_+(T)\equiv\average{(R-T)^+}=\int_{T}^{\infty} (R-T) p(R)dR.
\label{mu+def}
\end{equation}
Note that from these definitions we can obtain the following expression
\begin{equation}
\mu_+(T)-\mu_-(T)=\mu-T=\lambda\sigma.
\label{mu+-}
\end{equation}
where we also take into account the definition of $\lambda$ given by Eq.~(\ref{lambda}).

We have already obtained the $\mu_+(T)$ average in the case that the returns are Gaussian. From Eqs.~(\ref{mu+G}) and~(\ref{mu+-}), we thus have
\begin{equation}
\frac{\mu_-(T)}{\sigma}=-\lambda N(-\lambda)+N'(\lambda),
\label{mu-G}
\end{equation}
and
$$
\frac{\mu_+(T)}{\sigma}=\lambda N(\lambda)+N'(\lambda).
$$
Therefore, the Gain-Loss Ratio reads
\begin{equation}
\mbox{GL}(T)=\frac{\mu_+(T)}{\mu_-(T)}=1+\frac{\lambda}{N'(\lambda)-\lambda N(-\lambda)}.
\label{glG}
\end{equation}
Note that no arbitrage corresponds to $\lambda=0$ which means that average $\mu$ equals the target return $T$.
Also observe that the GL has not time units. This means that the annual GL should have the same value as the monthly GL. This is a very interesting and powerful property that avoids any discussion about the way we derive the annualized risk indicator as it happens with the Sortino ratios and the Adjusted Ratio Sharpe.

The bounds of the ratio are
\begin{equation}
0\leq\mbox{GL}(T)\leq\infty,
\label{glbounds}
\end{equation}
which respectively correspond to $T\rightarrow \infty$ and $T\rightarrow -\infty$. The GL, at least in the Gaussian framework, is a non decreasing function in terms of lambda whose fastest growing correspond to $\lambda>0$ regime, that is $T<\mu$. Thus, this risk measure is very sensitive to small changes when $\lambda>0$ while for negative lambda the ratio does not provide too much information. 

Figure~\ref{gl-fig} confirms this very last statement. The same plot also depicts the empirical results for a broad range of annualized target returns between $-30\%$ and $+30\%$. The more pronounced Gaussian behaviour and in a broader domain of lambda again corresponds to the ``Equity Market Neutral" strategy although other styles such as the ``Managed Futures" also follows nicely the curve. In contrast with the SoR and UPR risk measures, it is much more difficult to detect groups with similar values. Therefore, in this sense, previous measures seem to be more appropriate than the Gain-Loss Ratio.

We could also observe that the Gaussian GL gives a good measure for moderate values of positive lambdas ($0<\lambda<1$) to an important number of indices. In fact, within this regime, there are only two trading styles below the Gaussian performance. This behaviour appears to be also in clear contrast with the previous risk indicators.

\section{The Omega function\label{omega-sec}}

There exists another risk measure which may represent a different way of evaluating the Gain-Loss ratio. Their authors do not tell anything about this fact~\cite{keating,keating1}. The GL ratio authors are mainly worried about the benchmark risk-adjusted probability measure while the so-called Omega function authors consider that their indicator can be looked simply as another representation of the probability distribution of the underlying. This is certainly a different perspective but we here show that the Omega function is equivalent to the GL under quite general conditions. As to the case of the Gain-Loss ratio, we do not look at the fundamentals of the economic theory that lies behind but simply focuss on the statistical properties of the downside averages we take.

Keating and Shadwick~\cite{keating,keating1} and proposes the following Omega function measure:
\begin{equation}
\Omega(T)=\frac{I_2(T)}{I_1(T)},
\label{omega}
\end{equation}
where
$$
I_2(T)=\int_{T}^\infty (1-F(R))dR \qquad \mbox{and} \qquad I_1(T)=\int_{-\infty}^T F(R)dR
$$
using that $F(R)$ is the cumulative distribution function of the return $R$, i.e.,
$$
F(R)=\int_{-\infty}^R p(x)dx.
$$

We now try to evaluate the expressions for $I_1$ and $I_2$. Firstly, integrating by parts we have
\begin{equation}
I_1(T)=\int_{-\infty}^T F(R)dR=TF(T)-\lim_{R\rightarrow-\infty} RF(R)-\int_{-\infty}^TRp(R)dR.
\label{I1}
\end{equation}

Before proceeding further we show how the second summand is zero under some circumstances. We note that 
$$
\lim_{R\rightarrow-\infty} RF(R)=\lim_{R\rightarrow-\infty} R^2p(R).
$$
The limit value will thus depend on the way the pdf $p(R)$ decays to 0 as $R\rightarrow-\infty$. As far as $p(R)$ decays faster than $1/R^2$, this second summand can be neglected. This will be for instance the case of the Gaussian and the Laplace distributions or even a power law with a tail index larger than 2. However, the next steps of these calculations would not be applicable to the power law distributions with a slower decay, that is: $p(R)\sim 1/|R|^\alpha$ with $1<\alpha\leq 2$ as $R\rightarrow-\infty$. In this latter case, all the  moments of the pdf are infinite.

We now come back to Eq.~(\ref{I1}). Observe that it is also possible to rewrite the expression as
$$
TF(T)-\int_{-\infty}^TRp(R)dR=\int_{-\infty}^T (T-R) p(R)dR,
$$
and finally see (cf. Eq.~(\ref{mu-def})) that
\begin{equation}
I_1(T)=\average{(T-R)^+}=\mu_-(T).
\label{i1}
\end{equation}
Secondly, we can do the same with $I_2$. Under the condition that the pdf decays faster than $1/R^2$ as $R\rightarrow \infty$, similar calculations apply and lead us to state that (cf. Eq.~(\ref{mu+def}))
\begin{equation}
I_2(T)=\int_{T}^\infty (1-F(R))dR=\mu_+(T).
\label{i2}
\end{equation}

Therefore, according to the values derived for $I_1$ and $I_2$ and the definition given by Eq.~(\ref{gl}), we find that the Omega function and the Gain-Loss Ratio coincide since
\begin{equation}
\Omega(T)=\frac{\mu_+(T)}{\mu_-(T)}=\mbox{GL}(T).
\label{equal}
\end{equation}
We must however insist that this is only true when the pdf asymptotically decays faster than $1/R^2$ when $|R|\rightarrow \infty$. In such a case, the Omega will have the same bounds and behavior as the GL and results in Section~\ref{g-lsec} can be also applied to the Omega function.
In favour of $\Omega$ indicator, we may say that its way to handle the empirical data is more reliable (less noisy) specially when we have a small number of points. We have left to Appendix~\ref{omegarisk} some other new results related to what Keating and Shadwick call the "Omega risk".

\section{Error analysis: A discussion\label{errors}}

\begin{figure}
\begin{center}
\includegraphics{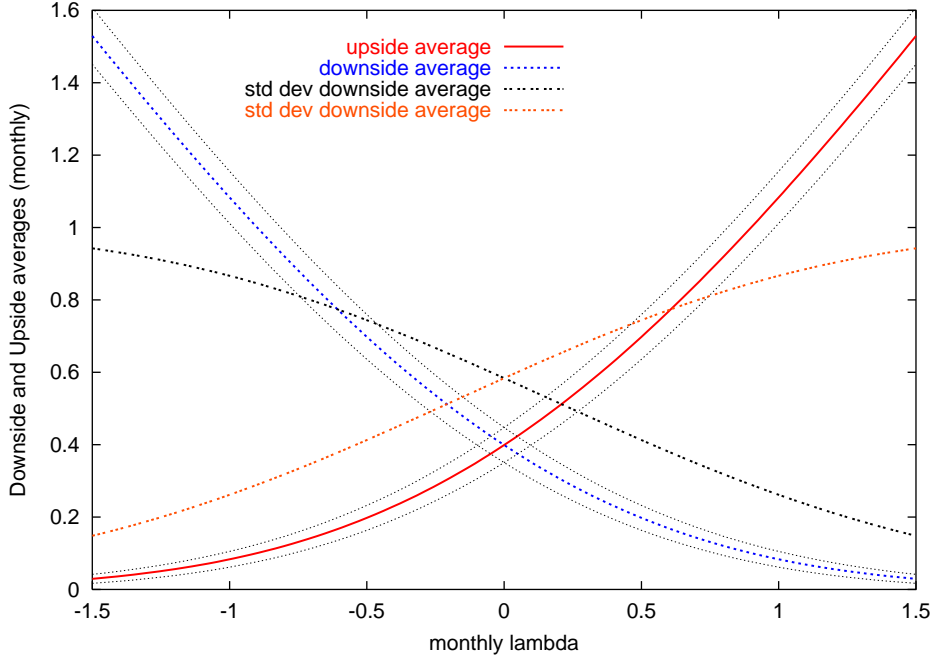}
\end{center}
\caption{The standard error behaviour for the Downside averages when underlying is Gaussian distributed and normalized with respect to $\sigma$. Upside and downside averages (solid lines) are accompanied with the standard error (dashed lines) for $M=145$ as given by Eqs.~(\ref{d11})--(\ref{d22}). The standard deviations of the same estimators only adopts values between 1 and 0. Upside and Downside averages have an opposite behaviour with respect to lambda.}
\label{theoretical-fig}
\end{figure}

\begin{figure}
\begin{center}
\includegraphics{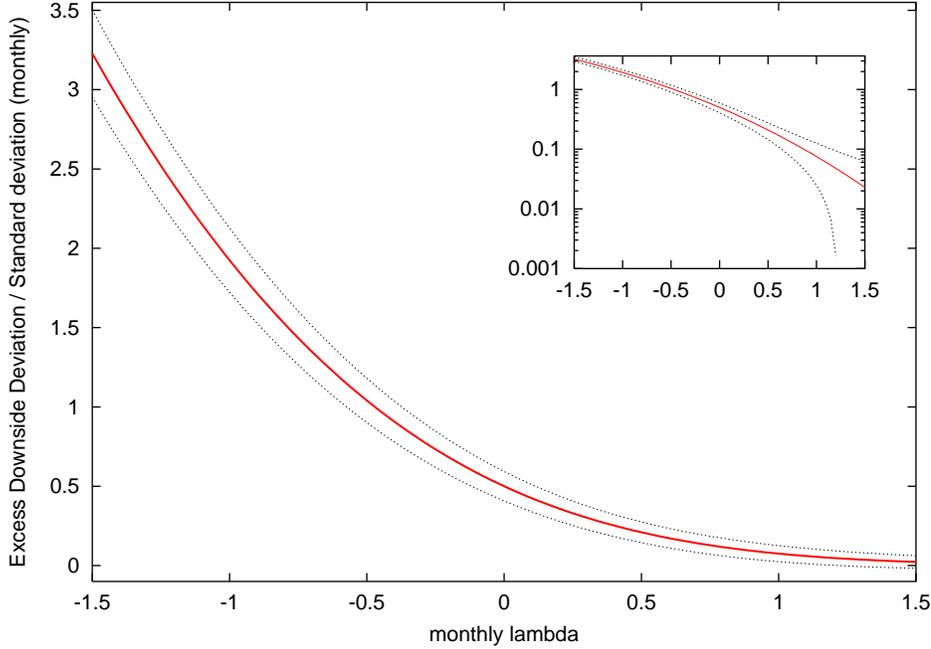}
\end{center}
\caption{The standard error behaviour for the Excess Downside Deviations when underlying is Gaussian distributed and normalized with respect to $\sigma$.}
\label{theoretical-fig1}
\end{figure}

The error analysis of the Downside Risk averages is a quite unexplored territory. We do not have here the aim of making a deep analysis on this topic since this deserves a whole paper. We would however like to provide at least few insights under the perspective of the Hedge Funds universe we have here studied.

Under the hypothesis that underlying follows a Brownian motion (returns are Gaussian distributed) we can quantify the error of the upside averages, the downside average and the excess downside deviation. The easiest thing to do is to assign its error magnitude to the standard deviation of those estimators. In such a case, one can obtain 
\begin{equation}
\frac{1}{\sigma^2}E \left[{(T-R)^+}^2 \right]=\frac{d(T)^2}{\sigma^2}=(\lambda^2+1)N(-\lambda)-\lambda N'(\lambda)
\label{d1}
\end{equation}
and 
\begin{equation}
\frac{1}{\sigma^2}E \left[{(R-T)^+}^2 \right]=(\lambda^2+1)N(-\lambda)+\lambda N'(\lambda).
\label{d2}
\end{equation}
If we join these expressions with the upside and downside averages given by Eqs.~(\ref{mu+G}) and~(\ref{mu-G}), we can derive the standard deviations and subsequently the standard errors: 
\begin{eqnarray}
\mbox{StdErr}\left[(T-R)^+\right]=\frac{1}{\sqrt{M-1}}\left\{E \left[{(T-R)^+}^2 \right]-\mu_-(T)^2\right\}^{1/2},\label{d11}
\\
\mbox{StdErr}\left[(R-T)^+\right]=\frac{1}{\sqrt{M-1}}\left\{E \left[{(R-T)^+}^2 \right]-\mu_+(T)^2\right\}^{1/2},\label{d22}
\end{eqnarray}
where $M$ is the data points used for the estimation.
The result of these operations is also given in Fig.~\ref{theoretical-fig}.
One should notice that the values of these functions are constrained between 0 and $1/\sqrt{M-1}$ and their behaviours are quite similar being antisymetric with respect to lambda. We can observe that when we take $M=145$ which corresponds to our Hedge Fund data set length the relative error is always below the 15\%. For the EDD case, the calculation is a bit longer but one can also obtain
\begin{equation}
\frac{1}{\sigma^4}E \left[{(T-R)^+}^4 \right]=(\lambda^4+6\lambda^2 +3)N(-\lambda)-(\lambda^3+4\lambda) N'(\lambda).
\label{d3}
\end{equation}
Following the same procedure as to the one for the upside and downside average, we can finally obtain its standard error. The aspect of its behaviour both in a linear and logarithmic scales is provided in Fig.~\ref{theoretical-fig1}. We can observe that the relative error is again below 15\% in almost the whole regime of lambda where we get the empirical observations, that is: $-1.5<\lambda<1$. In absolute values the error increases as lambda becomes more negative but its relative value might be quite high for $\lambda>1$.

\begin{figure}
\begin{center}
\includegraphics{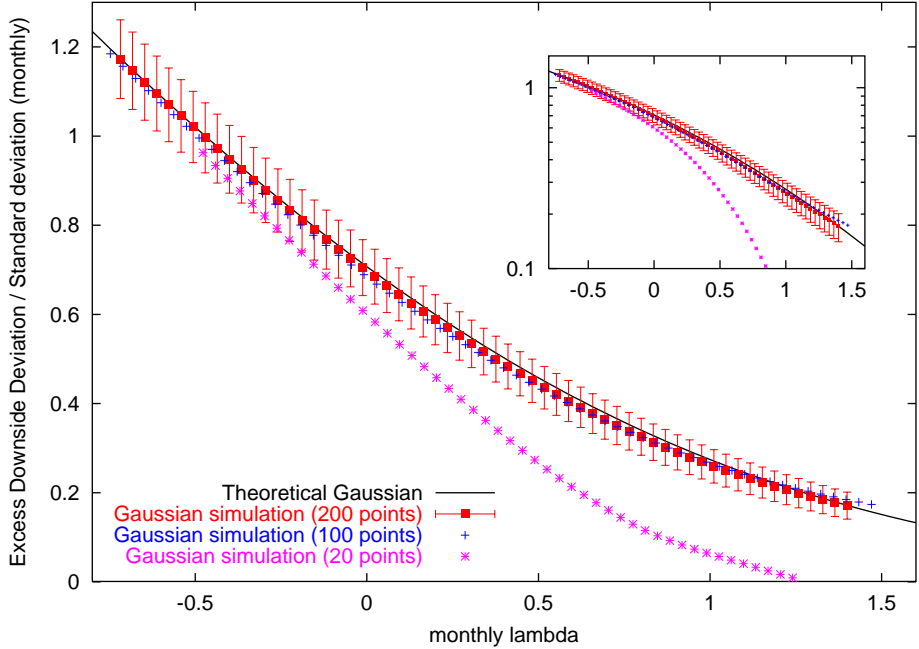}
\end{center}
\caption{The Excess Downside Deviation for a Gaussian simulated data set in terms of the monthly lambda. We present the theoretical curve when underlying is Gaussian distributed and compare it with simulations of different lenghts. The convergence to theoretical values is quite fast.}
\label{simulation}
\end{figure}

The moderate values for this relative errors certainly let us believe that the risk measures provided in the previous Sections are reliable enough. We must however take these bars to the assignation of the errors of these magnitudes with real care since these estimators are strongly biassed as one can see from the definitions in Section~\ref{dr}. The standard deviation would solely provide the order of magnitude of that error. A more accurate analysis of the empirical probability density we have behind each Hedge Fund style would be required to get a more reliable estimation of the degree of confidence of our observations.

We can go a bit further in this discussion and also check the convergence of these estimators. We have simulated a Bownian realization for the monthly returns with the empirical mean and volatility of the CST Index shown in Table~\ref{tab}. Having 20, 100 and 200 simulated timesteps, we have computed the Excess Downside Deviation in the same way as in previous sections. Figure~\ref{simulation} shows that the convergence to the theoretical curve is quite fast. With 100 steps we already have a reasonable estimation of the EDD and with a larger data set what we mainly achieve is a reduction of the error bars of our estimation. We have done the same procedure with the upside and downside averages obtaining similar conclusions. Therefore, this experiment also seems to confirm that the results provided in previous sections are quite reliable. At least under the Gaussian assumption, the convergence of our estimators is fairly reached for the number of points empirically available. 

\section{Final remarks\label{conc}}

Hedge Funds have enjoyed increasing levels of popularity coupled with opacity and some myths~\cite{lhabitant,seco}. We here have followed this recent interest in studying the Hedge Funds for an academic purpose (see for instance Refs.~\cite{susinno,nish} published in this journal). This is possible since data such as the Credit Suisse/Tremont Investable Hedge Fund Index is now easily available.

The strong non Gaussian character of financial markets have led to consider risk measures alternatives to CAPM theory in the context of the Downside Risk~\cite{lhabitant,till2,till,estrada}. The measures are able to distinguish between good and bad returns compared to our own personal target $T$ in a very simple manner. 
While the CAPM theory takes the average return growth and the return variance, the Downside Risk framework uses another two statistical measures which indeed keep folded some information of higher order moments. In particular, we have focused on the following risk measures: the Adjusted Ratio Sharpe, the Sortino ratios and the Gain-Loss Ratio from both a theoretical and empirical points of view. We have seen that the Downside Risk framework provides quite robust measurements and it appears to be the most natural extension to the CAPM theory and its mean-variance framework. 

The Hedge Funds is a field where these risk measures have most promising future. There are mainly two reasons. The first reason is the existence of wild fluctuations and pronounced negative skewness in data. And secondly is that there are few empirical data points available (of the order of hundreds of points). This last reason makes impossible to work with other more sophisticated risk metrics which are more sensitive to the wildest fluctuations. 

However, we have also seen that the Gaussian results for the studied Downside Risk measures are still important. We have shown that they work very well as a benchmark if we represent the empirical risk measures in terms of a modified Ratio Sharpe $\lambda=(\mu-T)/\sigma$. Perhaps quite surprisingly, we can also see in Figs.~\ref{sortino-fig},~\ref{upr-fig}, and~\ref{gl-fig} that the Gaussian trading investment behavior works better than most of the sophisticated trading style indices. The main reason lies on the fact that a Hedge Fund provides high benefits with the cost of having in most cases a negative skewness. Downside Risk measures take into account this asymmetry and include it to the risk perception. This is therefore another argument for using the Downside framework since Ratio Sharpe might wrongly overvalue the quality of a Hedge Fund by ignoring the skewness (and the kurtosis) effects in risk analysis.

To summarize, let us point out the main new contributions of the current paper. We have aimed to revisit some of the most popular Downside Risk indicators and have gone further on their application to the Hedge Fund universe in two different aspects. Firstly, the use of the same data set has enabled to get a reliable comparison between these risk indicators which is certainly difficult to get through the existing literature. And secondly, we have derived new analytical expressions for these risk indicators in terms of $\lambda=(\mu-T)/\sigma$ that allow to get a closer idea on the exact meaning of each indicator. In the same lines, we want to stress the fact that the modified Ratio Sharpe $\lambda$ appears to be a more useful parameter to work with instead of the target return $T$. It becomes very helpful to put the results in a broader context and in reference to a benchmark distribution such as the Gaussian one or even to compare different trading styles indices. Thanks to this fact, we have detected several indices with quite similar behaviour for the Sortino ratios and in the quotient $d(T)/\sigma$. However, the Gain-Loss Ratio is blind to this structure and unable to untangle these trading styles. For all these reasons and others discussed above, we can conclude that among the Sortino ratios the UPR is the best risk indicator for the Hedge Fund universe. We should however take all these conclusions with real care due to the small data set available. There is a strong presence of noise that we have here quantified with the standard error and the subsequent error bars computed with error propagation. To check the soundness of our conclusions we have discussed the error in the Downside Risk Metrics when underlying is Gaussian distributed. We have observed that the relative error based on our empirical data set length is quite reasonable being below the 15\%. We have also shown via simulations that the convergence of our estimators is quite fast since a hundred points is enough to get a reliable estimation of the Gaussian theoretical values involved in the Downside Risk Metrics.

There are many other interesting aspects to study under the current perspective. One possibility is to deeper study these risk indicators when returns follow another return distribution like a Laplace~\cite{schmidhuber,schmidhuber1} or for instance a power law distribution. We could also compute the here presented risk measures when target return is another asset. Another possibility is to study the effect of these analysis in the multi-factor market modeling~\cite{markowitz,cochrane,cochrane1,cochrane2,low}. In any case, all these topics should be left for future investigations.

\ack

The author acknowledges support from Direcci\'on General de Investigaci\'on under contract FIS2006-05204.

\appendix

\section{The Omega risk\label{omegarisk}}

The authors~\cite{keating,keating1} also define the Omega risk as
\begin{equation}
\Omega_R(T)\equiv\frac{\partial \ln \Omega}{\partial T}=\frac{1}{\Omega}\frac{\partial \Omega}{\partial T}.
\label{omrisk}
\end{equation}
This variable wants to measure the sensitivity of the Omega function to the changes in the target return $T$. Therefore, according to the definition~(\ref{omega}),
$$
\Omega_R(T)=\frac{1}{I_2}\frac{\partial I_2}{\partial T}-\frac{1}{I_1}\frac{\partial I_1}{\partial T}.
$$
But, from Eqs.~(\ref{i1})--(\ref{i2}) and taking into account that~\cite{footnote}
$$
\frac{\partial I_2}{\partial T}=F(T)-1 \qquad \frac{\partial I_1}{\partial T}=F(T),
$$
we finally obtain
$$
\Omega_R(T)=-\left[\frac{1}{I_2}+\left(\frac{1}{I_1}-\frac{1}{I_2}\right)F(T)\right].
$$

Assuming that the returns are Gaussian and recalling that $\lambda=(\mu-T)/\sigma$, we have that
$$
\Omega_R(T)=\left(\frac{1}{\mu_-}-\frac{1}{\mu_+}\right)N(\lambda)-\frac{1}{\mu_-},
$$
since $\mu_-$ is given by Eq.~(\ref{mu-G}), $\mu_+$ is given by Eq.~(\ref{mu+G}),
$$
\frac{\partial \mu_-}{\partial \lambda}=-\sigma N(-\lambda)
\qquad \mbox{and} \qquad
\frac{\partial \mu_+}{\partial \lambda}=\sigma N(\lambda).
$$
The Omega risk is always negative since the $\Omega$ is a non-decreasing function. Finally, we should mention that under an affine transformation of the form
$$
T\rightarrow\phi(T)=AT+B
$$
the function transforms as
$$
\Omega\rightarrow
\left\{
\begin{array}{cc}
\hat{\Omega}\left[\phi(T)\right]=\Omega(T) & \mbox{if } A>0;\\
\hat{\Omega}\left[\phi(T)\right]=1/\Omega(T) & \mbox{if } A<0.
\end{array}
\right.
$$
This is also true for the Gain-Loss Ratio equivalent risk indicator.


\begin{thebibliography}{999}
\bibitem{footnote} Authors of the Finance Development Centre give these expressions. These can be different depending on the calculus convention taken.
\end{thebibliography}
\end{document}